\documentstyle[prd,aps,epsf,graphicx,amsmath,amsthm,amssymb,psfrag]{revtex}

\newcommand{\half}{\mbox{$\textstyle \frac{1}{2}$}}
\newcommand{\thalf}{\mbox{$\textstyle \frac{3}{2}$}}

\newcommand{\tquat}{\mbox{$\textstyle \frac{3}{4}$}}

\begin{document} 

\tighten
\draft
\preprint{DAMTP-1999-??}
\twocolumn[\hsize\textwidth\columnwidth\hsize\csname 
@twocolumnfalse\endcsname

\title{Entanglement Induced Phase Transitions} 

\author{
Dorje C. Brody$^{*}$, Lane P. Hughston$^{\dagger}$ and 
Matthew F. Parry$^{\sharp}$
} 

\address{* $\sharp$ The Blackett Laboratory, Imperial College, 
London SW7 2BZ, UK} 

\address{$\dagger$ Department of Mathematics, King's College 
London, The Strand, London WC2R 2LS, UK} 

\date{\today} 

\maketitle 

\begin{abstract}
Starting from the canonical ensemble over the space of pure quantum
states, we obtain an integral representation for the partition function.
This is used to calculate the magnetisation of a system of $N$
spin-$\frac{1}{2}$ particles. The results suggest the existence of a new
type of first order phase transition that occurs at zero temperature in
the absence of spin-spin interactions. The transition arises as a
consequence of quantum entanglement. The effects of internal interactions
are analysed and the behaviour of the magnetic susceptibility for a small
number of interacting spins is determined. 
\end{abstract} 


\vskip2pc] 

In classical statistical mechanics, a phase transition is a phenomenon
characteristic of systems having internal interactions \cite{gold}.
Indeed, if the form of the internal energy is reasonably idealised, there
is a wide range of models describing phase transitions at finite
temperatures that can be solved exactly \cite{baxter}.  The situation is
similar for quantum systems, though exact solutions are typically
difficult to obtain in this case \cite{lieb}. 

One distinguishing feature of quantum systems in this context is the
existence of entanglement. This gives rise to a colossal increase in the
state space volume for combined systems. Furthermore quantum entanglement
can be viewed in itself as a form of internal interaction between the
constituent particles of the system. It is natural therefore to inquire
whether entanglement has any role to play in critical phenomena. 

In this letter we consider an entangled system of distinguishable
spin-$\frac{1}{2}$ particles, weakly interacting with a heat bath. A
general integral formula for the partition function is obtained and used
to analyse the magnetisation and the magnetic susceptibility of the
system. In particular we show that the magnetisation of $N$ noninteracting
particles in the low temperature limit behaves like
\begin{eqnarray}
M \sim \frac{1}{2} N \mu -
\left( 2^N-1 \right) \frac{k_B T}{ B},
\label{eq:15} 
\end{eqnarray}
where $\mu$ is the magnetic moment of an individual particle, $B$ is the
applied magnetic field, $T$ is the temperature of the heat bath, and $k_B$
is Boltzmann's constant. The $N$-dependence of the magnetisation points,
in the thermodynamic limit $N\!\rightarrow\!\infty$, to the possible
existence of a first order phase transition at zero temperature in the
absence of spin-spin interactions. This is a phenomenon with no classical
counterpart, indicating that entanglement may have an effect similar to
that of a mean-field which enhances critical phenomena. 

Let $\psi^{\alpha}$ be a state vector in a complex Hilbert space ${\cal
H}$. A physical observable, such as the Hamiltonian $H^{\alpha}_{\beta}$,
is represented by a linear operator acting on ${\cal H}$. If a system is
in the state $\psi^{\alpha}$, then the quantum expectation of an operator
$F^{\alpha}_{\beta}$ in that state is $\langle \hat{F}\rangle =
{\bar\psi}_{\alpha}F^{\alpha}_{\beta}
\psi^{\beta}/{\bar\psi}_{\gamma}\psi^{\gamma}$, where
${\bar\psi}_{\alpha}$ is the complex conjugate of $\psi^{\alpha}$. The
expectation $\langle \hat{F}\rangle$ is unaltered under transformations of
the form $\psi^{\alpha} \rightarrow \Lambda\psi^{\alpha}$, where $\Lambda$
is any nonvanishing complex number.  In other words, the Hilbert space
formulation of quantum mechanics carries an extra complex degree of
freedom given by the overall scale and phase. It is convenient, therefore,
to introduce the space of equivalence classes of vectors in ${\cal H}$
modulo complex scale transformations. This is the projective Hilbert space
${\cal PH}$: a point in ${\cal PH}$ corresponds to all the points on a ray
through the origin of ${\cal H}$, except the origin itself. We refer to
${\cal PH}$ as the space of pure quantum states. 

For definiteness, let us take ${\cal H}$ to be an $(n\!+\!1)$-dimensional
complex Hilbert space.  Then the corresponding state space is the
$n$-dimensional complex projective space ${\mathbb C}P^{n}$. The state
space ${\mathbb C}P^{n}$, when regarded as a real manifold ${\sl\Gamma}$
of dimension $2n$, is equipped with a natural symplectic structure, as
well as a Riemannian structure known as the Fubini-Study metric
\cite{cantoni}.  Each point $x$ in ${\sl\Gamma}$ represents a ray in
${\cal H}$.  Conversely, suppose we assign to each point $x$ in
${\sl\Gamma}$ a nonvanishing element $\psi^{\alpha}(x)$ in the ray
corresponding to $x$. Then given any observable $F$ acting on ${\cal H}$,
we can construct a real-valued biquadratic function $F(x)$ on
${\sl\Gamma}$ given by the expectation of $F^{\alpha}_{\beta}$ in the
state $\psi^{\alpha}(x)$. It is straightforward to verify that $F(x)$ is
independent of the particular choice of cross-section $\psi^{\alpha}(x)$. 

When we shift emphasis from ${\cal H}$ to ${\sl\Gamma}$, the role of the
complex number is taken over by the symplectic structure, given by a
nondegenerate skew-symmetric tensor field, and the Schr\"odinger equation
can be expressed in Hamiltonian form. The formulation of quantum mechanics
on the state space ${\sl\Gamma}$ thus bears a striking resemblance to
classical Hamiltonian mechanics, with the additional constraint that the
Hamiltonian $H(x)$ in quantum mechanics is of the special form $H(x) =
H^{\alpha}_{\beta}\Pi^{\beta}_{\alpha}(x)$, where
\begin{equation}
\Pi^{\beta}_{\alpha}(x) = \frac{{\bar\psi}_{\alpha}(x)
\psi^{\beta}(x)}{{\bar\psi}_{\gamma}(x)\psi^{\gamma}(x)}
\end{equation}
is the projection operator corresponding to the pure state
$x\!\in\!\Gamma$. 

The fact that the space ${\sl\Gamma}$ of pure states is the quantum phase
space makes ${\sl\Gamma}$ the appropriate space in which to consider the
effects of quantum entanglement. In this letter our interest is in the
weak coupling limit between a quantum system and an environment with which
it is in thermal equilibrium.  Specifically, we would like to know how to
characterise thermal equilibrium states on ${\sl\Gamma}$, that is, to find
an ensemble $\rho(x)$ over ${\sl\Gamma}$ that describes the distribution
of wave functions. This question has been addressed in \cite{brody}, where
it is shown that, if we assume the resulting distribution over
${\sl\Gamma}$ maximises the Shannon entropy associated with $\rho(x)$,
then the equilibrium ensemble is given by the Gibbs measure
\begin{eqnarray} 
\rho(x) = \frac{e^{-\beta H(x)}}{Z(\beta)} , 
\label{eq:5} 
\end{eqnarray} 
where $\beta=1/k_B T$, and
\begin{eqnarray} 
Z(\beta) = \int_{{\sl\Gamma}}e^{-\beta H(x)} dV 
\label{eq:6} 
\end{eqnarray} 
is the partition function. Here $dV$ is the volume element arising 
from the Fubini-Study metric on ${\sl\Gamma}$. More specifically, 
we have 
\begin{eqnarray} 
dV = \frac{D^{n}{\bar\psi}D^{n}\psi}{
({\bar\psi}_{\alpha}\psi^{\alpha})^{n+1}}  
\label{eq:7} 
\end{eqnarray} 
for the phase space volume element, where 
\begin{eqnarray} 
D^{n}\psi = \epsilon_{\alpha\beta\gamma\cdots\delta} 
\psi^{\alpha}d\psi^{\beta}d\psi^{\gamma}\cdots d\psi^{\delta} 
\label{eq:8} 
\end{eqnarray} 
and $\epsilon_{\alpha\beta\gamma\cdots\delta}$ is the totally 
skew-symmetric tensor. 

The interpretation of the distribution (\ref{eq:5}) is as follows. For
each pure state $x\!\in\!{\sl\Gamma}$ we compute the expectation $H(x)$ of
the Hamiltonian operator, conditioned on that pure state. For thermal
equilibrium the probability that the quantum system is in the pure state
$x$ is determined by the Boltzmann weight (\ref{eq:5}). Then the
unconditional expectation of the Hamiltonian in the thermal state is given
by the internal energy
\begin{eqnarray} 
U(\beta) = \int_{{\sl\Gamma}} H(x)\rho(x)dV ,
\label{eq:9} 
\end{eqnarray} 
which is equivalent to the trace formula $H^{\alpha}_{\beta}
\rho_{\alpha}^{\beta}$, where $\rho_{\alpha}^{\beta}$ is the density
matrix associated with the distribution (\ref{eq:5}), given by
\begin{equation}
\rho_{\alpha}^{\beta} = \int_{{\sl\Gamma}} \rho(x) 
\Pi_{\alpha}^{\beta}(x) dV.
\end{equation}
It follows by a standard identity that $U=-\partial\ln Z/\partial\beta$.
Therefore, we would like to obtain an explicit expression for the
partition function $Z(\beta)$, in order to determine properties of the
thermodynamic functions. 

For the computation of the partition function, it is convenient to revert
to the homogeneous coordinates $\psi^{\alpha}$ ($\alpha=0,1,\cdots,n$) on
the complex projective space ${\mathbb C}P^{n}$. Then, we observe that the
integration in (\ref{eq:6}) can be lifted to ${\mathbb C}^{n+1}$ with a
spherical constraint ${\bar\psi}_{\alpha}\psi^{\alpha}=1$, which gives us
\begin{eqnarray} 
Z(\beta) = \int_{{\mathbb C}^{n+1}}\delta(\bar{\psi}\psi-1) 
e^{-\beta H^{\alpha}_{\beta}{\bar\psi}_{\alpha}\psi^{\beta}} 
d^{n+1}{\bar\psi} d^{n+1}\psi . 
\label{eq:10} 
\end{eqnarray} 
Moreover, if we choose a basis such that the Hamiltonian operator is
diagonal and substitute the standard identity $\delta(x) = (2\pi)^{-1}
\int e^{{\rm i}\xi x}d\xi$ for the $\delta$-function, then the ${\mathbb
C}^{n+1}$-integration becomes a Gaussian and we obtain
\begin{eqnarray} 
Z(\beta) = \int_{-\infty}^{\infty}
\frac{(2\pi)^{n}e^{-{\rm i}\xi}d\xi}
{(\beta E_{0}-{\rm i}\xi)(\beta 
E_{1}-{\rm i}\xi)\cdots (\beta E_{n}-{\rm i}\xi)}, 
\label{eq:11} 
\end{eqnarray} 
where $E_{k}$ are the energy eigenvalues. In obtaining (\ref{eq:11}) we
transform from complex coordinates $\psi^{\alpha}$ to real coordinates,
which makes the integration involved for each energy eigenvalue $E_{k}$ in
(\ref{eq:10}) a double Gaussian. If we analytically continue $\xi$ to the
lower half-plane, then the integrand decreases exponentially and we can
close the contour and apply the residue theorem to evaluate the integral.
The result is
\begin{eqnarray}
Z(\beta) = \sum_{\mbox{\scriptsize residues}} e^{\lambda} \left(
\prod_{k=0}^{n}\frac{1}{\lambda+\beta E_{k}}\right) ,
\label{eq:12}
\end{eqnarray}
where we have written $\lambda=-{\rm i}\xi$ and we have discarded the
physically unimportant factors of $2\pi$. 

Formula (\ref{eq:12}) for the partition function is valid for any quantum
system that can be modelled by a finite dimensional state space. In
particular, if there is no degeneracy in the energy eigenvalues, then we
find that the partition function reduces to the following simple
expression: 
\begin{eqnarray} 
Z(\beta) = \sum_{k=0}^{n}e^{-\beta E_{k}} \left( 
\prod_{l=0,\neq k}^{n}\frac{1}{\beta(E_{l}-E_{k})}\right) .  
\label{eq:13} 
\end{eqnarray} 
This formula is applicable, for example, to the case of a single
spin-$\frac{n}{2}$ particle in a magnetic field, for which we can write
$E_{k}=-(\frac{n}{2}-k)\mu B$, where $k=0, \cdots, n$.  Then we have
\begin{equation}
Z=\frac{1}{n!}\left(\frac{\sinh{\half\beta\mu B}}{\half\beta\mu 
B}\right)^{n}
\end{equation}
for the partition function. Interestingly, this is identical to the
expression one obtains for $n$ classically indistinguishable, independent
(disentangled) spin-$\frac{1}{2}$ particles in a magnetic field. In the
spin-$\frac{1}{2}$ case (i.e. $n\!=\!1$), as was addressed in
\cite{brody}, the magnetisation energy is given by
\begin{equation}
U=k_{B}T-\frac{1}{2}\mu B\coth\half\beta\mu B.
\end{equation}

Let us now apply the result (\ref{eq:12}) to a system of distinguishable
spin-$\frac{1}{2}$ particles (e.g., electrons on a lattice) in a magnetic
field. Classically, if the particles are not interacting, then the
resulting partition function factors, and we obtain the same magnetisation
per particle as in the single particle case. However, quantum
mechanically, this is no longer the case, because of the existence of
quantum entanglement. That is, even in the absence of direct interactions,
the presence of entanglement implies that the particles are not
independent. For example, if the system consists of two spin-$\frac{1}{2}$
particles, any entangled state (such as the singlet state) gives rise to a
nonvanishing Boltzmann weight through formula (\ref{eq:5}). 

For $N$ noninteracting spin-$\frac{1}{2}$ particles, the Hamiltonian 
operator is 
\begin{eqnarray} 
{\hat H} = - \mu B \sum_{i=1}^{N} {\hat s}_{iz} = - \mu B {\hat S}_{z},
\label{eq:14a} 
\end{eqnarray} 
where ${\hat{\bf S}}=\sum_{i=1}^{N}{\hat{\bf s}}_{i}$. Note that, although
the total number of eigenstates is given by $2^N$, the corresponding
eigenvalues are highly degenerate. In particular, there are only $N\!+\!1$
distinct energy eigenvalues, given by $\epsilon_{k} = -\left(
\frac{N}{2}-k \right)\mu B$, where the index $k$ runs from $0$ to $N$. The
degree of degeneracy associated with the eigenvalue $\epsilon_{k}$ is
${}_{N}C_{k}$.  Using this energy spectrum, we have determined the
magnetisation of the system per particle for $N$ up to five, with the
resulting plot given in Fig. 1. 

The results in Fig. 1 show that, even in the absence of spin-spin
interactions, the existence of quantum entanglement gives rise to
nontrivial thermal expectation values---a result that has no classical
counterpart. Furthermore, it also indicates that, in the thermodynamic
limit $N\rightarrow\infty$, the gradient of the magnetisation at zero
temperature diverges. If this were the case, then we would obtain a first
order phase transition at $T=0$ such that the value of magnetisation would
jump from $0$ to $1/2$ as $T\downarrow0$.  However, because the formula
(\ref{eq:11}) is not guaranteed to be valid in the thermodynamic limit, we
cannot prove the existence of a phase transition. Nevertheless, we may
consider the asymptotic behaviour of the magnetisation in the low
temperature limit, and study how the resulting expression depends on $N$.
In particular, as $T\rightarrow0$ the contribution from the residue
arising from the ground state energy dominates the partition function. 
Thus we find
\begin{eqnarray}
Z &\sim& e^{-\beta \epsilon_0} \prod_{k=1}^{N} \frac{1}{[\beta(
\epsilon_k-\epsilon_0)]^{{}_N C_k}} \nonumber \\
&=& a_N (\beta\mu B)^{1-2^N}\exp{(\half N\beta\mu B)},
\end{eqnarray}
where $a_N$ is independent of $\beta$.  Then using $M
=\beta^{-1}\partial\ln Z/\partial B$, we obtain the behaviour of the
magnetisation as given in (\ref{eq:15}). This result is valid for an
arbitrarily large but finite $N$, in the limit $T\rightarrow0$, and agrees
with the numerical results in Fig. 1. In particular, the coefficient of
$T$ diverges exponentially in $N$, strongly pointing towards the existence
of a transition. 

\begin{figure}[thb] 
\label{fig1} 
\psfrag{y}[tb]{$\frac{M}{N\mu}$}
\psfrag{x}[bt]{\scriptsize$k_B T/\mu B$}
\psfrag{N1}[bl]{\scriptsize$N\!=\!1$}
\psfrag{N2}[bl]{\scriptsize$2$}
\psfrag{N3}[bl]{\scriptsize$3$}
\psfrag{N4}[bl]{\scriptsize$4$}
\psfrag{N5}[bl]{\scriptsize$5$}
\includegraphics[width=8cm,angle=-90]{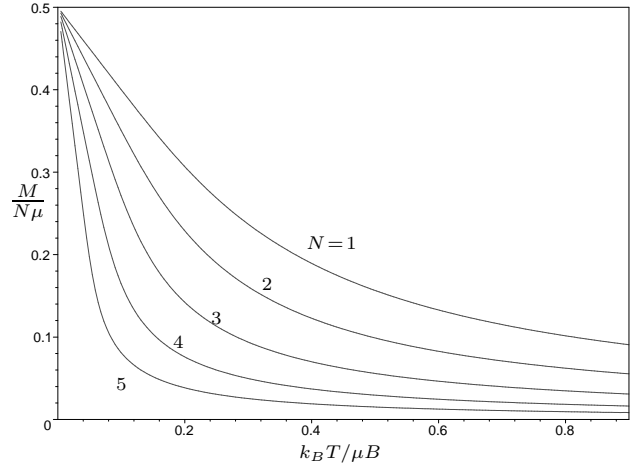} 
\vspace{-1cm}
\caption{The magnetisation per particle for noninteracting 
spin-$\frac{1}{2}$ 
particles. Classically, the result is independent of the number 
$N$ of particles. Quantum mechanically, as $N$ is increased, the 
effect of entanglement accumulates to give quantitatively 
different behaviours for different $N$.}
\end{figure} 

The phase transition implied by the foregoing analysis is, of course,
rather artificial, in the sense that the transition takes place at zero
temperature, and that the magnetisation can take only two distinct values.
This is not surprising however because the model we have considered has no
spin-spin interactions. If we include interactions in order to have a
finite temperature transition, then we expect the decrease of $M/N$ with
increasing $N$ at fixed $T$, as observed in Fig. 1, is suppressed. 

We now turn to study a highly interacting system in which derivations of
thermodynamic functions are analytically tractable. In particular, we
consider the quantum Ising model on a complete graph, rather than the
conventional square-lattice models. Therefore, the Hamiltonian operator is
now given by
\begin{eqnarray} 
{\hat H} &=& - \mu B \sum_{i=1}^{N} {\hat s}_{iz} 
-J \sum_{i>j}{\hat{\bf s}}_{i}\cdot {\hat{\bf s}}_{j} \nonumber \\ 
&=& - \mu B {\hat S}_{z} - \half J ({\hat S}^{2}-\tquat N) , 
\label{eq:16} 
\end{eqnarray} 
Without loss of generality, we can assume that the total number $N=2n$ of 
the spins is even. Then (\ref{eq:12}) gives us 
\begin{equation} 
Z(\beta) = \sum_{\mbox{\scriptsize residues}} 
e^{\lambda} \left( \prod_{s=0}^{n}\prod_{m=-s}^{s}
\frac{1}{(\lambda+\beta \epsilon_{m}(s))^{d_{2n}(s)}}\right), 
\label{eq:17} 
\end{equation} 
where
\begin{equation}
\epsilon_{m}(s)=-m\mu B -\frac{1}{2} J\left(s(s+1)-\thalf n\right)
\end{equation}
and
$d_{2n}(s)={}_{2n}C_{n-s}-{}_{2n}C_{n-s-1}$. The ground state 
energy, in particular, is given by $E_{0}=\epsilon_{n}(n)$. 

With the expression (\ref{eq:17}) at hand, we can consider the behaviour
of thermodynamic functions. In particular, for a system of interacting
spins, we can determine the second moment, namely, the magnetic
susceptibility of the system. Using the definition $\chi = (1/N)\,\partial
M/\partial B$ for the magnetic susceptibility, we analysed numerically the
behaviour of $\chi$ for a range of values of $N$. The results are shown in
Fig. 2, indicating the vanishing of the second moment at $T=0$ and
$T\rightarrow\infty$, as well as a peak at finite temperature, the
sharpness of which increases with $N$. 

\begin{figure}[htb] 
\label{fig2} 
\psfrag{y}[tb]{$\frac{\chi}{\mu/B}$}
\psfrag{x}[bt]{\scriptsize$k_B T/\mu B$}
\psfrag{N1}[bc]{\scriptsize$N\!=\!1$}
\psfrag{N2}[bl]{\scriptsize$2$}
\psfrag{N3}[bl]{\scriptsize$3$}
\psfrag{N4}[bl]{\scriptsize$4$}
\psfrag{N5}[bl]{\scriptsize$5$}
\includegraphics[width=8cm,angle=-90]{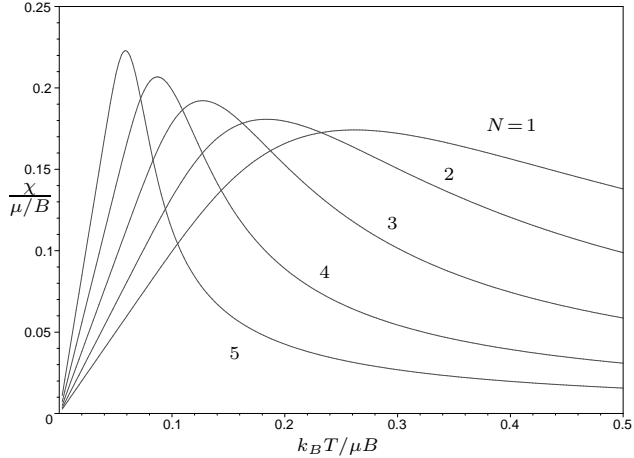} 
\vspace{-1cm}
\caption{Magnetic susceptibility of the system of interacting 
spin-$\frac{1}{2}$ particles on a complete graph for $N=1,\cdots,5$. 
The numerical value $J/\mu B = 0.2$ has been chosen for the 
interaction strength. } 
\end{figure} 

There are two regimes of (\ref{eq:17}) which can be explored analytically.
The first is the case of weak coupling in the low temperature limit. More
precisely, when $J/\mu B \leq 2/N$, the residue coming from the ground
state energy is again the leading contribution to the partition function
as $T\rightarrow 0$. If one calculates the magnetisation per particle,
then, as conjectured, the decrease of $M/N$ with increasing $N$ at fixed
$T$ is suppressed. However, the analytic result obtained is of limited use
for large $N$, because the restriction on $J$ means one is simultaneously
taking $J$ to zero. 

The second regime is the strong coupling limit $J\rightarrow\infty$. In
this limit, the system is effectively quenched, so we expect to observe
little impact from quantum entanglement. When we calculate the partition
function, we find, for $J/\mu B \gg 1$, that the main contribution to $Z$
is due to the residues arising from the $2n\!+\!1$ states that have $s=n$.
Furthermore, the energy difference between states of differing $s$ can be
treated as independent of $\mu B$.  Specifically we have
\begin{eqnarray}
Z &\sim& \sum_{k=-n}^n e^{-\beta \epsilon_k(n)}\left( 
\prod_{s=0}^{n-1}\prod_{m=-s}^s
\frac{1}{[\beta(\epsilon_m(s)-\epsilon_k(n))]^{d_{2n}(s)}}\right. 
\nonumber \\
&{}&\hskip1.5cm\times \left. \prod_{l=-n,\neq k}^n  
\frac{1}{\beta(\epsilon_l(n)-\epsilon_k(n))}\right)\nonumber \\
&\simeq& \frac{e^{\frac{1}{2}\beta J n (n-\frac{1}{2})}}{N!\, \beta^{2^N-1} 
(\mu B)^N} \sum_{k=-n}^n {}_N C_{n-k}\, (-1)^{n-k} e^{k\beta\mu 
B}\nonumber \\
&=& a_N(J)\frac{e^{\frac{1}{8}\beta J N (N-1)}}{\beta^{2^N-1}(\mu B)^N} 
\sinh^N{(\half \beta \mu B)}.
\end{eqnarray}
It is now straightforward to determine the magnetic susceptibility:
\begin{equation}
\frac{\chi}{\mu/B} = \left[1 - \left(\frac{\frac{1}{2} \beta \mu 
B}{\sinh{\frac{1}{2} \beta \mu B}} \right)^2\:\right] \frac{k_B T}{\mu B}.
\end{equation}
That this result is independent of $N$ is not surprising since we are
considering interacting spins on a complete graph in the strong coupling
limit. The effect of quantum entanglement is thus washed out in this
limit. 

In summary, we have obtained an expression for the partition function of a
finite quantum system in thermal equilibrium, when the equilibrium states
are obtained by maximising the Shannon entropy on the space of pure
states. We applied this result to study the thermal expectation of the
magnetisation and magnetic susceptibility of a system of $N$
spin-$\frac{1}{2}$ particles. In the case of noninteracting spins, we were
able to determine explicitly the effect arising from quantum entanglement.
In reality, individual particles forming magnetic substances are not
likely to be fully entangled. Nevertheless, there are partial
entanglements within a system, and entanglements appear to have the effect
of enhancing phase transitions. 

DCB gratefully acknowledges financial support from The Royal 
Society. 

$*$ Electronic mail: {\tt dorje@ic.ac.uk }
 
$\dagger$ Electronic mail: {\tt lane.hughston@kcl.ac.uk }

$\sharp$ Electronic mail: {\tt mparry@ic.ac.uk }

\end{document}